\documentclass[preprint,12pt]{elsarticle}

\usepackage{amssymb}
\usepackage{dcolumn,amsmath}
\usepackage{bm}
\usepackage{multirow}

\usepackage{soul}
\usepackage{amsfonts}
\usepackage{cancel}

\usepackage{color}

\makeatletter
\def\ps@pprintTitle{%
  \let\@oddhead\@empty
  \let\@evenhead\@empty
  \let\@oddfoot\@empty
  \let\@evenfoot\@oddfoot
}
\makeatother

\begin{document}

\begin{frontmatter}

\title{Axion-mediated electron-electron interaction in RaOCH$_3$ molecule}

\author[label1,label2]{Anna Zakharova}
\ead{zakharova.annet@gmail.com}
\ead{anna.zakharova@spbu.ru}

\author[label1]{Mikhail Reiter}
\ead{mikh.reiter@gmail.com}

\affiliation[label1]{organization={St. Petersburg State University},
addressline={7/9 Universitetskaya nab.}, 
city={St. Petersburg},
postcode={199034}, 
state={},
country={Russia}}

\affiliation[label2]{organization={Petersburg Nuclear Physics Institute named by B.P. Konstantinov of National Research Centre
"Kurchatov Institute"},
addressline={1, mkr. Orlova roshcha}, 
city={Gatchina},
postcode={188300}, 
country={Russia}}
\begin{abstract}

We study the parity-violating electron-electron interaction mediated by the axion-like in the hexatomic molecule of a symmetric top type. The rich rovibrational behavior require electronic computations for multiple molecular configurations which can be reduced using Generalized Relativistic Effective Core Potential. To restore the correct behavior in the core region we use a one-center restoration technique generalized by us earlier to the two-electron properties. The property is averaged on the lowest-lying rovibrational states.

\end{abstract}


\end{frontmatter}

\section{Introduction}
\label{intro}

The axion, which is the pseudoscalar particle with a nonzero mass, is believed~\cite{cosmo:1983, abbott1983cosmological,dine1983not, chadha2022axion} to satisfy the requirements for being the constituent of the Dark matter. Proposed as a pseudo-Nambu-Goldstone boson arising from the spontaneous breaking of Peccei-Quinn symmetry~\cite{PhysRevLett.38.1440}, it is expected~\cite{peccei1977cp} to solve the strong CP problem in the QCD.
Thus, observing the axion would be a significant step toward solving a number of problems in modern theoretical physics.


The study of molecular properties allows one to investigate effects associated with the violation of spatial parity ($\mathcal{P}$) and time-reversal ($\mathcal{T}$) symmetries. The most well-known and studied example is the electric dipole moment of the electron (eEDM)~\cite{PhysRevLett.126.171301}. Generally speaking, other interactions such as scalar-pseudoscalar electron-nucleon interactions (NE-SPS) mimic eEDM impact. However, the sensitivities of different molecules and different states may vary, which may help to distinguish these effects. E.g. this may be helped by the dependence of the property on the configuration of the molecule. While the local effective operators corresponding to eEDM and NE-SPS have weak dependence~\cite{zakharova2022rotating, zakharova2021rovibrational, zakharova2024symmetric, zakharova2022impact}, it may be stronger for the long-range interactions mediated by light particles, such as electron-nucleon and electron-electron interactions mediated by the axion~\cite{maison2021axion, maison2021electronic, prosnyak2024axion}.

The symmetric-top molecules appear promising as a new platform for parity violation searches, thanks to the closely-spaced opposite-parity $l$- and $K$-doublets \cite{hutzler2020polyatomic}.
Certain symmetric-top molecules with heavy elements, e.g., RaOCH$_3$ and YbOCH$_3$, allow laser-cooling that enables precision measurements. Moreover, these measurements can be performed within the ground-electronic state, and states without transverse vibrations of nuclear bonds can be isolated. As a result, the study of such molecules is simplified from both experimental and theoretical points of view. The rovibrational nature of these states may result in the different sensitivities of the different levels to the long-range effects.

The extensive electronic computations for multiple molecular configurations required for averaging over the rovibrational wavefunction may be greatly simplified in the approach of the Generalized Relativistic Effective Core Potential (GRECP). Regretfully, GRECP computations spoil the behavior of the wavefunction in the core region, which is highly important for the properties of interest. This may be fixed by the one-center restoration technique \cite{titov1999generalized, titov2006d, Petrov:02}. Recently, we have generalized this computation for the two-electron axion-mediated interaction \cite{new_el_el_axion}. In this work, we use this technique to study the impact of the electron-electron axion-mediated interaction on the RaOCH$_3$ molecule.



\newpage

\section{Axion-mediated electron-electron interaction in molecules}

Consider the interaction of axions, described by the pseudoscalar field $a$, with the leptons of the Standard Model $\Psi_f$. It may include $\mathcal{P}$, $\mathcal{T}$-even scalar interaction, as well as a $\mathcal{P}$, $\mathcal{T}$-odd pseudoscalar terms:
\begin{equation}
\mathcal{S}_{\rm int} = a\sum_{f\in\mathrm{SM}} \bar{\Psi}_f\Big(g_{f,S} + i g_{f,P}\gamma_5\Big) \Psi_f.
\end{equation}

The tree axion-exchange process results in the long-range $\mathcal{P}$, $\mathcal{T}$-odd interaction with the Yukawa-type potential, see, e.g., \cite{PRD:1984:130, PRB:2017:127}. The exponent of the Yukawa potential can be approximated as $e^{-m_a r} \sim 1$, since the typical molecular length scale corresponds to an axion mass of $\sim 1$ KeV.  Then, in the coordinate representation, axion-mediated electron-electron interaction is expressed through two-electron integrals of the form:
\begin{equation}
A(a,b|c,d) = \int d \vec{r_1} d \vec{r_2} \;\psi_a^\dagger(\vec{r_1}) (i \gamma_0 \gamma_5) \psi_c (\vec{r_1}) \; \frac{1}{r_{12}} \; \psi_b^\dagger(\vec{r_2}) \gamma_0 \psi_d (\vec{r_2}),
\label{eq:ax_two_el_int}
\end{equation}
where $r_{12} = |\vec{r_1}-\vec{r_2}|$.
Note that the Coulomb interaction of electrons is described by a similar expression without $\gamma_0$ and $i \gamma_0 \gamma_5$:
\begin{equation}
C(a,b|c,d) = \int d \vec{r_1} d \vec{r_2} \;\psi_a^\dagger(\vec{r_1}) \psi_c (\vec{r_1}) \; \frac{1}{r_{12}} \; \psi_b^\dagger(\vec{r_2}) \psi_d (\vec{r_2}).
\label{eq:c_two_el_int}
\end{equation}
With the two-electron integrals of the form~(\ref{eq:ax_two_el_int}) in mind, one can construct the electron-electron interaction potential:
\begin{equation}
V_{ee}(\vec{r}_1, \vec{r}_2) = \frac{g_{e,S} g_{e,P}}{4 \pi r_{12}} (\gamma_0)_1 (i\gamma_0 \gamma_5)_2, \end{equation}
where the subscript $()_i$ indicates which electron the gamma matrix acts on. 

To estimate the sensitivity of the considered molecule to the axion detection, we introduce the so-called enhancement parameter $E_{\rm ax}$. The enhancement parameter is the cornerstone of this study, since, e.g., the energy shift of opposite-$K$ levels (where $K$ is the quantum number for the projection of the total angular momentum on the molecular axis) in an external electric field will be proportional to:
\begin{equation}
E_{+M} - E_{-M} = P \cdot g_{e,S} \cdot g_{e,p} \cdot E_{\rm ax}.
\end{equation}
Here the coefficient $P$ is the degree of polarization, that may be set to $1$ if the electric field is strong enough, see Ref.~\cite{petrov2022sensitivity}.

The enhancement parameter is defined in the following manner
\begin{equation}
E_{\rm ax} = \frac{\langle \Psi_{\rm el} | \hat{H}_{ee} | \Psi_{\rm el} \rangle}{\Omega \cdot g_{e, S} \cdot g_{e, P}}.
\label{eq:enh_param}
\end{equation}
Here $\Omega$ is the projection of the
total electronic angular momentum on the molecular axis and $H_{ee}$ is the Hamiltonian of $\mathcal{P}$, $\mathcal{T}$-odd interaction, defined as:
\begin{equation}
\hat{H}_{ee} = \sum_{\substack{i,j=1 \\ i \neq j}}^{N_e} V_{ee}(\vec{r}_i, \vec{r}_j),
\end{equation}
where the summation is over all electrons of the system. In the Eq.~(\ref{eq:enh_param}), $\Psi_{\rm el}$ is the electron wave function. It depends on the geometry of the molecule, so after calculation of this parameter one should average it over rovibrational nuclear wavefunction.

In the Kramers-restricted SCF approximation the molecular orbitals are grouped into the closed-shell pairs related by $\mathcal{T}$ transform. Also for the open-shell molecule one orbital which we will denote as $n$ is unpaired. Then the property averaged over the electronic wavefunction may be represented as \cite{new_el_el_axion},
\begin{equation}
\langle \Psi_{el}|\hat{H}_{ee}|\Psi_{el}\rangle = \sum_{m\in {\text closed shells}} \Big( A(nm|nm) - A(nm|mn) - A(mn|nm)\Big)
\end{equation}

We refer the reader to our paper \cite{new_el_el_axion} for the details on the electronic wavefunction computations using the one-center restoration technique.

\section{Geometry of the molecule}

\begin{figure}
\centering
\includegraphics[width=0.5\linewidth]{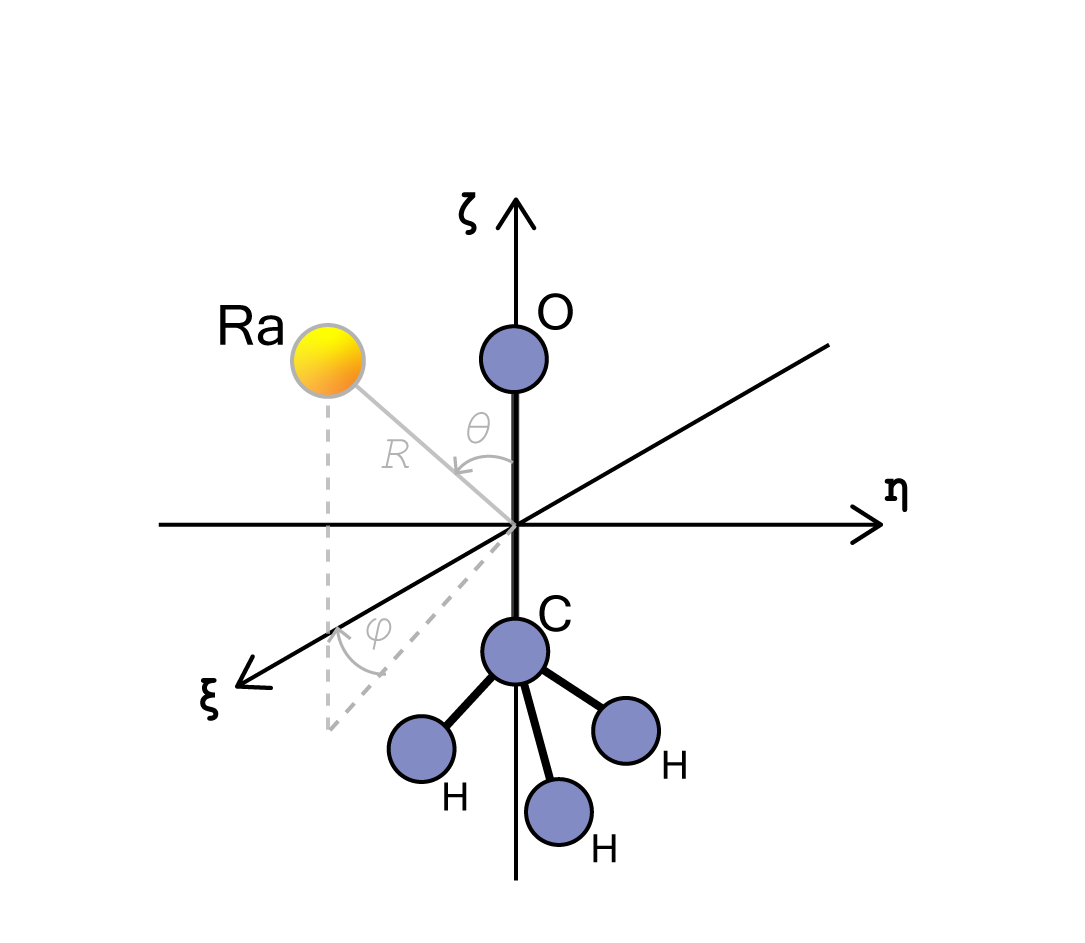}
\caption{The RaOCH$_3$ molecule (new picture...)
\label{fig:raoch3}}
\end{figure}

The geometry of the molecule is determined by the geometry of the OCH$_3$ ligand and the position of the radium atom relative to the ligand. The ligand is considered rigid due to the fact that its bond stretching and bending frequencies are much higher than those of the Ra -- OCH$_3$ bond as in Ref.~\cite{zakharova2022rotating}. We use same ligand is  described by $r(O-C) =2.6\,\mathrm{a.u.}$, $R(C-H)=2.053\,\mathrm{a.u.}$, $\angle(O-C-H) =110.73^\circ$. The Ra position relative to the ligand is described by the vector $\vec{R}$, Ra -- ligand c.m., in the coordinate system associated with the ligand. It is specified by the length and two angles -- $\{R, \theta, \phi\}$, see Fig.~\ref{fig:raoch3}.
The position of a molecule of a fixed configuration in space is determined by the direction of the $\zeta$ axis, denoted in the space-fixed frame as $\vec{r}$, and the angle of rotation around it $\gamma$.

Within the Born-Oppenheimer approximation, the molecular wave function is the product of the electronic and nuclear wave functions:
\begin{equation}
\Psi_{\rm total} \sim \Psi_{\rm nuc}(R, \theta, \phi, \hat{r}, \gamma) \Psi_{\rm elec}(\{r_i\}| R, \theta, \phi).
\label{eq:B_O_appr}
\end{equation}
The dependence of $\Psi_{\rm elec}$ is due to the fact that the electronic wave function is uniquely determined by the configuration of the molecule; in order to describe the rovibrational structure, $\Psi_{\rm nuc}$ must be the function of position of the molecule in space.

\section{Rovibrational wave functions}

Once the electron wave functions are obtained and the enhancement parameter is calculated, the rovibrational wave function $\Psi_{\rm nuc}$ from Eq.~(\ref{eq:B_O_appr}) can be found. It satisfies the Schr\"{o}dinger equation of the form:
\begin{equation}
\hat{H}_{\rm nuc} \Psi_{\rm nuc}(R, \theta, \phi, \hat{r}, \gamma) = E \Psi_{\rm nuc}(R, \theta, \phi, \hat{r}, \gamma)
\label{eq:Shreq}
\end{equation}
with the adiabatic Hamiltonian describing nuclear motion. Since the OCH$_3$ ligand is considered rigid, the Ra - ligand center of mass system can be described as a linear rotor of length $R$. Such a rotor has an angular momentum $\hat{\vec{l}}$ in the space-fixed frame. Ligand, on the other hand, is considered rigid symmetric top rotor with moments of inertia $I_\zeta$ and $I_\eta = I_\xi$. Its angular momentum is $\hat{\vec{j}}$ with projections on the molecular axis $\{\hat{j_\xi}, \hat{j_\zeta}, \hat{j_\eta}\}$. The adiabatic nuclear Hamiltonian reads:
\begin{equation}
\hat{H}_{\rm nuc}=\left[-\frac{1}{2\mu}\frac{\partial^2}{\partial R^2}+\frac{\hat{\vec{l}}^2}{2\mu R^2}\right]+\left[ \frac{\hat{j_\xi}^2}{2 I_\xi} + \frac{\hat{j_\zeta}^2}{2 I_\zeta} + \frac{\hat{j_\eta}^2}{2 I_\eta}\right]+V(R, \theta, \phi, \hat{r}, \gamma),
\label{eq:nuc_ham}
\end{equation}
where $\mu$ is the reduced mass of the Ra-OCH$_3$ system, and $V$ is the adiabatic electron potential. To solve the equation~(\ref{eq:Shreq}) the coupled-channel technique developed for symmetric top molecules in is used\cite{zakharova2022rotating, zakharova2024symmetric}. The solution is decomposed into the eigenfunctions
of the $\hat{{j^2}}$, $\hat{{l^2}}$, and $\hat{{j_\zeta}}$:
\begin{equation}
\begin{gathered}
\Psi_{\rm nuc}(R, \theta, \phi, \hat{r}, \gamma) \equiv \Psi^{JM}(R, \theta, \phi, \hat{r}, \gamma) = \\ = \sum_{j=0}^{j_{\rm max}} \sum_{k=-j}^{+j} \sum_{l=0}^{l_{\rm max}} F_{jkl}^J(R) Y_{jkl}^{JM}(\theta, \phi, \hat{r}, \gamma),
\label{psiexp_RaOH}
\end{gathered}
\end{equation}
where $J$ and $M$ are the quantum numbers for the total angular momentum of the molecule and its projection on space-fixed axis, $j$ is the ligand momentum in the space-fixed frame and $k$ is its projection on the $\zeta$ axis in the ligand-fixed frame. For the numerical computations, infinite summations are truncated by parameters with the subscript "max". Angular-dependent part of $\Psi_{\rm nuc}$ is:
\begin{equation}
Y_{jkl}^{JM}(\theta, \phi, \hat{r}, \gamma) = \sum_{m=-j}^{+j} \sum_{m_l=-l}^{+l} \langle jmlm_l |JM \rangle Y_{l m_l}(\theta, \phi) D_{jmk} (\hat{r}, \gamma),
\end{equation}
where $\langle jmlm_l |JM \rangle$ is the Clebsh-Gordan coefficient, and the $D_{jmk}$ is the Wigner function. The projection of $\vec{j}$ on the space-fixed axis is $m$, and the projection of $\vec{l}$ is $m_l$. Then one can write the radial part as the Schr\"{o}dinger-like equation:
\begin{equation}
\begin{gathered}
\left(\frac{d^2}{d R^2} - \frac{l(l+1)}{R^2} + 2 \mu E - 2 \mu E_{\rm lig}^{jk} \right) F^J_{jkl}(R) = \\ = 2 \mu \sum_{j'k'l'} v_{jklj'k'l'}(R) F^J_{j'k'l'}(R).
\end{gathered}
\end{equation}
Here the $E_{\rm lig}^{jk}$ is the ligand Hamiltonian eigenstates:
\begin{equation}
E_{\rm lig}^{jk} = \frac{1}{2 I_\xi}j(j + 1) + (\frac{1}{2 I_\zeta} - \frac{1}{2 I_\xi}) k^2,
\end{equation}
and the matrix potential is
\begin{equation}
\begin{gathered}
{v}_{jkl\tilde{j}\tilde{k}\tilde{l}}(R)=(-1)^{j+\tilde{j}+\tilde{k}-J}\sqrt{\frac{[j][\tilde{j}][l][\tilde{l}]}{4 \pi}} 
\times \\ \times 
\sum_{\lambda \mu} \sqrt{[\lambda]} V_{\lambda \mu}(R) \begin{Bmatrix}j&l&J\\\tilde{l}&\tilde{j}&\lambda\end{Bmatrix}
\begin{pmatrix}l&\lambda&\tilde{l}\\0&0&0\end{pmatrix}\begin{pmatrix}j&\lambda&\tilde{j}\\-k&\mu&\tilde{k}\end{pmatrix}
\end{gathered}
\end{equation}
with the functions $V_{\lambda \mu} (R)$ obtained from the adiabatic potential expansion
\begin{equation}
V(R, \theta, \phi, \hat{r}) \equiv V(R,\theta, \phi) = \sum_{\lambda=0}^{\lambda_{\rm max}} \sum_{\mu = - \lambda}^{+\lambda} V_{\lambda \mu}(R) Y_{\lambda \mu} (\theta, \phi).
\end{equation}
The radial dependence is discretized using the eigenbasis of the longitudinal harmonic oscillators. This turns the problem into the diagonalization of the finite-dimensional sparse matrix \cite{zakharova2025}.

\section{Results and discussion}

The results for the lowest-lying rovibrational states are given in Tab.~\ref{tbl:table1}. One may notice that the sensitivity for the hexatomic molecule happens to be much stronger than for the triatomic ones.

\begin{table}[h]
\small
  \caption{Electron-electron axion-mediated interaction for RaOCH$_3$ molecule}
  \label{tbl:table1}
  \renewcommand{\arraystretch}{1.5}
  \begin{tabular*}{0.9\textwidth}{@{\extracolsep{\fill}}lllrr}
    \hline\hline
$v_\parallel$ & $v_\perp$ &  $l$ & $K$ &  $W_{ee}$,  $10^{-5}\lambda_e^{-1}$\\
\hline
  \multicolumn{5}{l}{Equilibrium configuration}\\
\hline
 &  &  & & 17.37\\
 \hline
  \multicolumn{5}{l}{Rovibrational states}\\
\hline
0 & 0 & 0 & 0 & 17.21\\
0 & 0 & 0 & $\pm 1$ & 17.21\\
0 & 1 & $\pm 1$ & 0 & 17.3\\
0 & 1 & $\pm 1$ & $\pm 1$ & 17.3\\
\hline\hline
  \multicolumn{5}{l}{The RaOH equilibrium configuration (SCF + GRECP + OCR) \cite{new_el_el_axion}}\\
  \hline
&  &  & & 1.383    \\

\hline\hline
  \multicolumn{5}{l}{The YbOH equilibrium configuration (FS-CCSD, $m_a=1..10$eV) \cite{maison2021electronic}}\\
  \hline
&  &  &  &  1.46   \\
 \hline\hline
  \multicolumn{5}{l}{The BaF equilibrium configuration (CCSD, $m_a=1..10^2$eV) \cite{prosnyak2024axion}}\\
  \hline
&  &  &  &  0.8   \\
 \hline\hline
  \end{tabular*}
\end{table}


\section{Acknowledgement}
The work was supported by the Russian Science Foundation (grant number 24-72-10060).

\bibliographystyle{apsrev}

\end{document}